\documentclass{elsart}
\usepackage{hyperref}
\usepackage{graphicx,times}
\usepackage{amssymb}
\hypersetup{
  pdfmenubar=true, pdftoolbar=true, pdfpagemode={None} }

\begin{document}

\begin{frontmatter}
\title{Dynamics of a piecewise smooth map with singularity}
\author[aloke]{Aloke Kumar}
\ead{alokek@gmail.com}
%\ead[url]{http://cc.domaindlx.com/aloke}
%\author[sb]{Soumitro Banerjee}
\author{Soumitro Banerjee \corauthref{cor1}\thanksref{sb}}
\corauth[cor1]{Corresponding author}
%\thanks[sb]{Corresponding author}
%\footnote{Corresponding author}
\ead{soumitro@ee.iitkgp.ernet.in}
%\ead[url]{http://www.ee.iitkgp.ernet.in/soumitro}
\author[dpl]{Daniel P. Lathrop}
\ead{dpl@complex.umd.edu}
\address[aloke]{Department of Mechanical
    Engineering,\\ Indian Institute of Technology, Kharagpur-721302,
    India}
\address[sb]{Department of Electrical 
    Engineering,\\ Indian Institute of Technology, Kharagpur-721302,
    India}
\address[dpl]{Department of Physics, and Institute of Physical
    Science and Technology,\\ University of Maryland, College Park, USA}
%\date{}

\begin{abstract}
Experiments observing the liquid surface in a vertically oscillating
container have indicated that modeling the dynamics of such systems
require maps that admit states at infinity. In this paper we investigate
the bifurcations in such a map. We show that though such maps in
general fall in the category of piecewise smooth maps, the mechanisms
of bifurcations are quite different from those in other piecewise
smooth maps. We obtain the conditions of occurrence of infinite
states, and show that periodic orbits containing such states are
superstable. We observe period-adding cascade in this system, and obtain
the scaling law of the successive periodic windows. 

\end{abstract}


\end{frontmatter}

\section{Introduction}

Recently a lot of research attention has been directed toward the
dynamics of piecewise smooth maps (PWS), because they represent a large
number of systems of practical interest including switching electrical
circuits and impacting mechanical systems. In such systems the
discrete phase space is divided into compartments within which the map
is smooth, and the compartments are separated by borderlines at which
the map is not differentiable. A one-dimensional piecewise smooth map has the general
form
\begin{equation}
x_{n+1} = f(x_n) = \left\{\begin{array}{cc} g(x_n,\mu), &
\mbox{for}\;\;\; x_n < \lambda\\
h(x_n,\mu), & \mbox{for}\;\;\; x_n > \lambda \end{array} \right.
\end{equation}
where $\mu$ is the bifurcation parameter and the compartments are
separated by the borderline value $\lambda$. In such a map, there is
the possibility that a fixed point may collide with the border with
the change of a system parameter. When that happens, there is  a
sudden change in the stability of the fixed point, leading to a new
type of nonlinear phenomenon called {\em border collision
  bifurcation} \cite{Nusse94,Nusse95}. It has been shown that such border collisions may lead
to atypical bifurcation phenomena like transition from period-2 to
period-3 or a sudden onset of chaos without undergoing the usual
period doubling cascade \cite{Nusse92}.

A few different forms of
such maps have been investigated to date:

\begin{enumerate}

\item The map $f$ is continuous, not differentiable at $\lambda$, and
both $dg/dx$ and $dh/dx$ are finite. Such maps represent a
class of switching circuits, and have been investigated in detail
\cite{Nusse95,1d98}.

\item The map $f$ is continuous, not differentiable at $\lambda$, and
there is a square root singularity (i.e.,  $dh/dx$ is
infinite) at one side of the border. Such maps represent the impact
oscillator \cite{nordmark,budd2,Nusse94}.

\item The map $f$ is discontinuous at $\lambda$, the derivative $df/dx$
is also discontinuous at $\lambda$, but the value of the derivative at
both sides of the border are finite. Such maps represent a class of
electronic circuits \cite{discontcircuit} including the Colpitts
oscillator \cite{maggio2000}, and the bifurcation theory
for such maps has been developed recently \cite{parag03}.

\end{enumerate}

\begin{figure}[tbh]
\centering \includegraphics[width=3in]{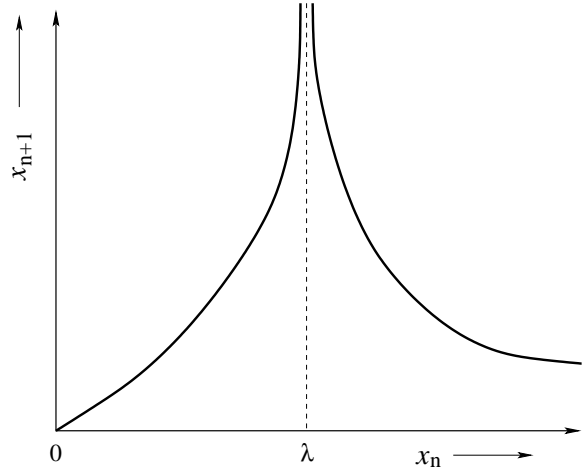}
\caption{The graph of the map given by (\ref{eqn:recast1}) and
(\ref{eqn:recast2}).}
\label{infmap}
\end{figure}

In 1997 an experiment was reported, where the oscillations in
the surface
of a liquid held in a vertically oscillating container were observed
using a laser probe, and it was found that under some conditions
narrow jets are ejected from the center of the surface. It was shown
that representation of this system required a map with not only slope
singularity but also magnitude singularity at the border \cite{WCGL97}.
The proposed map had the form

\begin{eqnarray}
 x_{n+1} &=&  \gamma x_n + \frac{\alpha
 x_n}{(x_n - \lambda)^2} \;\;\; \mbox{for}\;\;\; x_n<\lambda
 \label{eqn:recast1} \\
x_{n+1} &=& \beta + \frac{\rho x_n}{(x_n - \lambda)^2} \;\;\;\;\;\;\; \mbox{for}\;\;\;
 x_n> \lambda
 \label{eqn:recast2}
\end{eqnarray}
where $\alpha$, $\beta$, $\rho$ and $\lambda$ are constants and $\gamma$ is
the bifurcation parameter. The graph of the map is schematically shown
in Fig.~\ref{infmap}. 
%The investigators used a value of $\gamma \geq 1$
%due to the linear instability for the waves.
In this paper we investigate the bifurcation phenomena in a map of the
above form -- for which no theory is currently available.

\begin{figure}[tbh]
\centering  \includegraphics[width=4in]{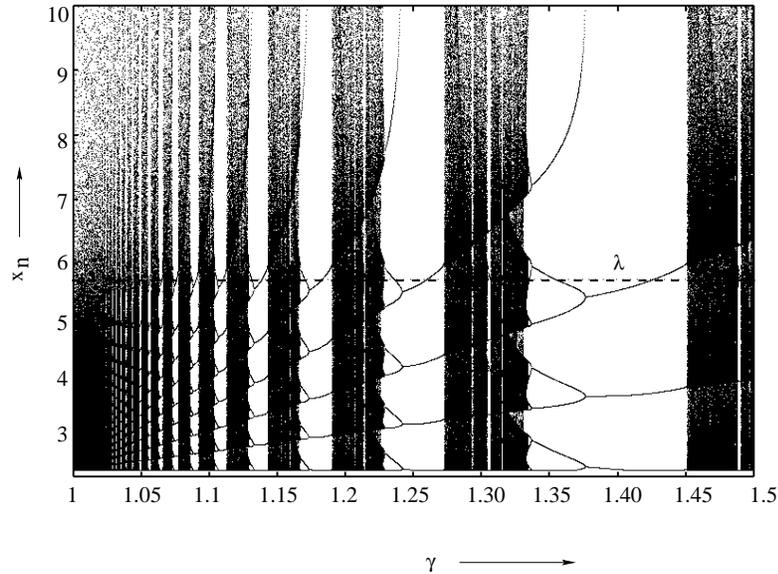}
  \caption{Bifurcation diagram with $\alpha=0.04$, $\beta=2.8$, $\rho=0.14$,
  $\lambda=5.76$, and $\gamma$ as the variable
  parameter. The diagram is truncated above $x_n = 10$.}\label{fig.bifurcation}
\end{figure}

In this map, the vertical line $x=\lambda$ forms an asymptote for
the equations (\ref{eqn:recast1}) and (\ref{eqn:recast2}), and a
singularity occurs at this value. This asymptotic behavior occurs
due to the geometric considerations in the waves --- that the
waveheight/wavelength cannot exceed some ratio before 
the wave becomes self-intersecting. While obtaining the
bifurcation diagram, if $x$ takes a value close to $\lambda$, the value of
$x$ at the next iterate is very high. The program has to account for this
possibility. The bifurcation diagram obtained this way is
presented in Fig.~\ref{fig.bifurcation}. As both domain and range
of the piecewise smooth map are represented by the half-open set $[0,\infty)$,
Fig.~\ref{fig.bifurcation} and all subsequent bifurcation diagrams
have been truncated.

A few features are noticeable in the bifurcation diagram. First,
there is a succession of periodic windows with the following properties:
\begin{enumerate}
\item Within each periodic window, as the bifurcation parameter
$\gamma$ is reduced, a period doubling cascade terminates in chaos.
\item If $n$ is the periodicity of the base orbit (lowest period orbit) at the highest parameter
value in a perodic window and $m$ is the periodicity of the base orbit at
the highest parameter value in the next window, then $m=n+1$,  i.e.,
there is a period adding cascade as the parameter is reduced.
\item The width of the periodic window reduces monotonically as the
period adding cascade progresses.
\end{enumerate}

\begin{figure}[tbh]
\centering  \includegraphics[width=4in]{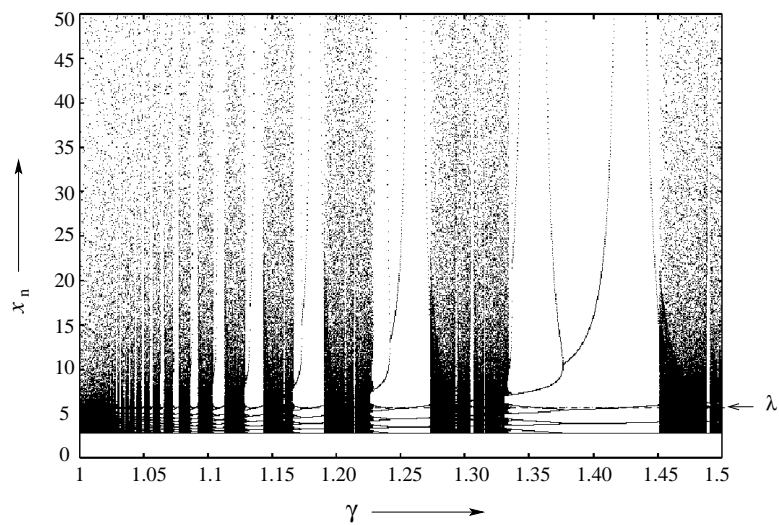}\\
  \caption{Bifurcation diagram with horizontal lines drawn at $ y = \beta$ and $y =\lambda$.}\label{fig.enlargedbif}
\end{figure}

In earlier studies, period adding cascades were observed in the
study of piecewise smooth maps of finite magnitude and finite slope
\cite{Nusse95,1d98} and those with square root singularity of
derivatives \cite{nordmark,budd2,Nusse94}.
In maps of the former type it has been found that each periodic
window originates at a border collsion. But for the system under
the present study, the
bifurcation diagram plotted for a larger range of $x$
(Fig.~\ref{fig.enlargedbif}) demonstrates that
each periodic window does not emerge due to border collision. These
occur at saddle-node bifurcations.

In systems with square
root singularity it was observed that a period-$n$ orbit always
has $n-1$ points on one side of the border and one point on the
other side. Such orbits were called maximal. Though from
Fig.~\ref{fig.bifurcation} this may seem true also for the system
presently under consideration, a closer scrutiny of
Fig.~\ref{fig.enlargedbif} shows that the orbits are not maximal
at the points of emergence of periodic windows. They become maximal
as the periodic point crosses the borderline value of $\lambda$.

\section{Border-collision bifurcations}
In view of the differences, therefore, the theory developed for
these other types of piecewise smooth maps  cannot be applied for the map with
magnitude singularity presently under consideration.

For this system, a few questions demand fresh answers. For which
parameter values does the state reach infinite value? It is easy
to see that in the chaotic windows, because of the ergodic nature
of the orbit, iterates come arbitrarily close to the value of
$\lambda$, and then the next iterate must assume infinite
value. Therefore, even though a bifurcation diagram cannot be
drawn for such high range of the variable, there must be points
at infinity for all parameter values in all the chaotic
windows. This is supported by the observation in \cite{WCGL97}
where such states were called ``ejecting states.''

This system reveals another interesting aspect: infinity is
reached even within periodic windows --- at the points of border
collision.

\begin{figure}[tbh]
\centering  \includegraphics[width=4in]{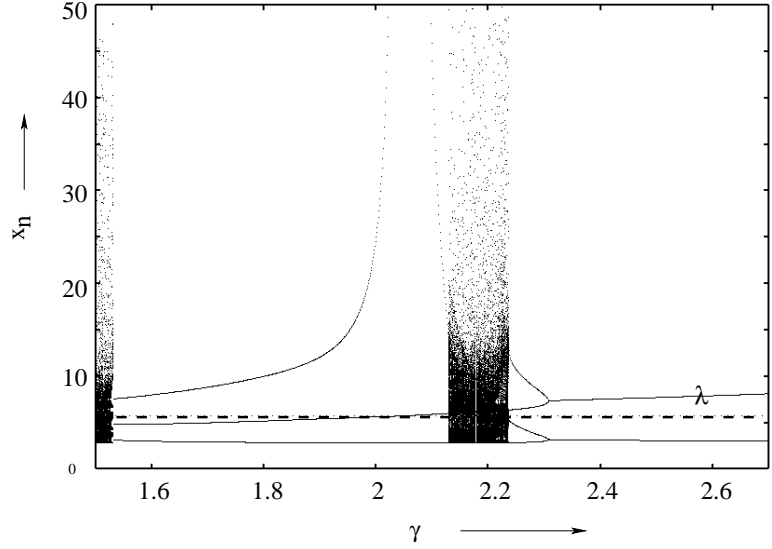}\\
  \caption{Close up of the period-3 window.}\label{fig.valid}
\end{figure}

 The points of border-collision in the periodic windows can be determined
theoretically. At the point of border collision, one point of the
periodic orbit must have the value $\lambda$. Since $\lambda \mapsto
\infty \mapsto \beta$, for  a period three orbit the three points are
$\beta,\lambda,\infty$. To find the value of the parameter $\gamma$ at the
crossover point we set
\begin{equation}\label{eqn:cross}
\gamma\beta + \frac{\alpha\beta}{(\beta-\lambda)^2} = \lambda
\end{equation}
This yields a value $\gamma = 2.052577$ for which the period-3 orbit
contains a point at infinity. Fig.~\ref{fig.valid} gives a closeup of
the period-3 window showing this point.

At the point where the period-4 orbit
touches infinity, three of the four points will be $\lambda,\infty,$
and $\beta$. The fourth point obtained by substituting $x_n =
\beta$ in (\ref{eqn:recast1}) is
\[ \gamma\beta + \frac{\alpha\beta}{(\beta-\lambda)^2}.\]
Since this point must map to $\lambda$, we obtain the parameter value
$\gamma$ from (\ref{eqn:recast1}) as
\begin{equation}
\gamma\{\gamma\beta + \frac{\alpha\beta}{(\beta-\lambda)^2}\}+
\frac{\alpha\{\gamma\beta +
\frac{\alpha\beta}{(\beta-\lambda)^2}\}}{(\gamma\beta +
\frac{\alpha\beta}{(\beta-\lambda)^2}-\lambda)^2} =
\lambda\label{eqn:firstlamda}
\end{equation}
This yields the following 4-th degree equation in $\gamma$
\begin{equation}\label{eqn:firstlamda1}
   a_0\gamma^4 +a_1\gamma^3 +a_2\gamma^2 +a_3\gamma +a_4 = 0
\end{equation}
where \[ a_0 = \beta^3\]
\[ a_1 = \frac{3\alpha\beta^3}{(\beta-\lambda)^2}-2\beta^2\lambda\]
\[a_2 =\frac{3\alpha^2\beta^3}{(\beta-\lambda)^4} - \frac{4\lambda\alpha\beta^2}{(\beta-\lambda)^2}+\beta\lambda^2-\beta^2\lambda\]
\[a_3 =\frac{\alpha^3\beta^3}{(\beta-\lambda)^6}-\frac{2\alpha^2\beta^2\lambda}{(\beta-\lambda)^4}+\frac{\alpha\beta\lambda^2}{(\beta-\lambda)^2}+\alpha\beta - \frac{2\alpha\beta^2\lambda}{(\beta-\lambda)^2}+2\beta\lambda^2\]
\[a_4 =\frac{\alpha^2\beta}{(\beta-\lambda)^2}-\frac{\alpha^2\beta^2\lambda}{(\beta-\lambda)^4}+\frac{2\lambda^2\alpha\beta}{(\beta-\lambda)^2}-\lambda^3.\]
Solving (\ref{eqn:firstlamda1})  using $\beta
=2.8,\lambda = 5.76,\alpha = 0.040$ yields $\gamma
=1.425510183648710 $. Out of the other three solutions, one is
negative and the other two complex conjugate --- which are not
possible values of the parameter.

%\begin{figure}[tbh]
%  % Requires \usepackage{graphicx}
%  \includegraphics[width=3.5in]{valid1}\\
%  \caption{Close up of bifurcation diagram at singularity point}\label{fig.valid1}
%\end{figure}
%A close scrutiny of the bifurcation diagram indicates that the map
%does indeed achieve singularity in the interval $[1.4,1.44]$ thus
%validating the result.

In a similar manner, all the periodic orbits with a point at infinity
and their corresponding parameter values can be obtained. The only
requirement for such border-collision points to exist is $\beta<\lambda$.

\section{Stability of periodic orbits}

The rate of expansion/contraction at any periodic point is dependent
on the derivative of the map function

\begin{eqnarray}
 \frac{d x_{n+1}}{d x_n} &=&
\gamma  + \frac{\alpha}{(x_n -
 \lambda)^2}- \frac{\alpha x_n}{(x_n -
 \lambda)^3} \;\;\;\;\; \mbox{for}\;\;\; x_n<\lambda ,
 \label{eqn:derivative1}\\
 \frac{d x_{n+1}}{d x_n} &=&
\frac{\rho}{(x_n - \lambda)^2} - \frac{2\rho x_n}{(x_n -
\lambda)^3} \;\;\;\;\;\;\;\;\;\;\; \mbox{for}\;\;\; x_n>\lambda .
 \label{eqn:derivative2}
\end{eqnarray}

Since border collision points are related to the singularity
condition, and since from (\ref{eqn:derivative1}) and
(\ref{eqn:derivative2}) we find that
\[ \lim_{x_n \rightarrow \lambda^-} \frac{d x_{n+1}}{d
x_n}\rightarrow \infty \;\;\;\mbox{ and }\;\;\; \lim_{x_n
\rightarrow \lambda^+} \frac{d x_{n+1}}{d x_n}\rightarrow -\infty,
\] one may tend to believe that periodic orbits containing a point
at infinity cannot be stable. However, since
\[\lim_{x_n \rightarrow \infty} \frac{dx_{n+1}}{dx_n}\rightarrow 0,\]
this need not be true, and so it is interesting to work out the
stability of the orbits containing infinity.

Such an orbit must have two points with values $\lambda$ and
$\infty$. The other points in the orbit have finite slope.
Therefore the derivative of $n$th iterate map for that
condition must be a finite number times the limiting values of
{\small
\begin{equation}
\left(\lim_{x_n \rightarrow \lambda^-} \gamma  + \frac{\alpha}{(x_n -
 \lambda)^2}- \frac{\alpha x_n}{(x_n -
 \lambda)^3}\right)\!\times \!\left( \lim_{x_{n+1}\rightarrow \infty} \frac{\rho}{(x_{n+1} - \lambda)^2} - \frac{2\rho x_{n+1}}{(x_{n+1} -
\lambda)^3}\right) \label{eqn:limit1}
\end{equation}}
and
{\small
\begin{equation}
\left(\lim_{x_n \rightarrow \lambda^+}\frac{\rho}{(x_n - \lambda)^2} -
\frac{2\rho x_n}{(x_n - \lambda)^3} \right)\!\times \!\left(
\lim_{x_{n+1}\rightarrow \infty} \frac{\rho}{(x_{n+1} -
\lambda)^2} - \frac{2\rho x_{n+1}}{(x_{n+1} - \lambda)^3}\right).
\label{eqn:limit2}
\end{equation}}
For obtaining the limiting values of (\ref{eqn:limit1}) and
(\ref{eqn:limit2}), we transform these into a limit of a single
expression. Using the symbolic computation facility of {\sc Matlab},
we substitute the expression for $x_{n+1}$, i.e., (\ref{eqn:recast1}) or
(\ref{eqn:recast2}) depending on the position of $x_n$.
For simplicity and compactness these
expressions are denoted by $F^-$ and $F^+$ as $x_n \rightarrow
\lambda^-$ or $x_n \rightarrow \lambda^+$. The limit expression
becomes
\begin{equation}
\lim_{x_n \rightarrow \lambda^-} F^- \label{eqn:simlim1}
\end{equation}
\begin{equation}
\lim_{x_n \rightarrow \lambda^+} F^+ \label{eqn:simlim2}
\end{equation}
We find that in both cases limit exists and is zero. Therefore we
conclude that the orbit including the point at infinity is
superstable. It should be
noted that the
value of limit is independent of $\beta,\lambda,\alpha,\rho$ and
hence the result is a general one.

The result was numerically confirmed by choosing a border
collision point and then by calculating the product of the
derivatives. It was observed that the product could be made
arbitrarily small by choosing appropriate value of the variable
$x$.

\section{Bifurcation behavior for $\beta\geq\lambda$}

\begin{figure}[tbh]
\centering  \includegraphics[width=4in]{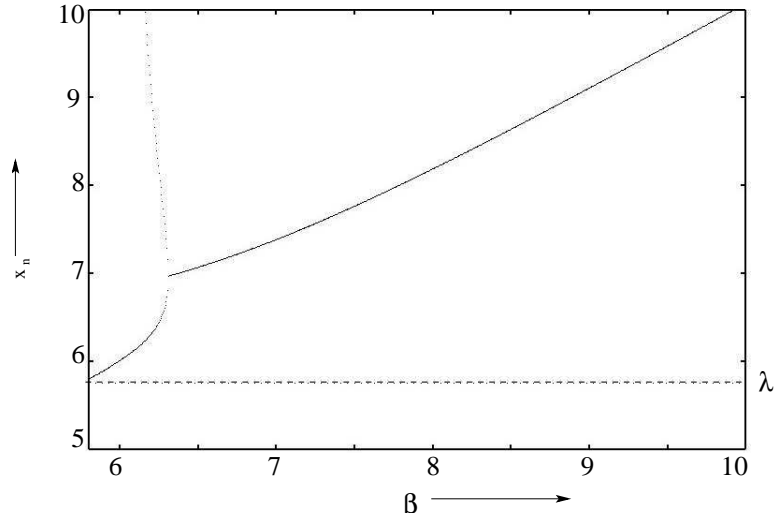}\\
  \caption{Bifurcation diagram with respect to $\beta$ for $\beta\geq\lambda$.}
  \label{betabifurcation}
\end{figure}

In the earlier  sections the bifurcation behavior for the case
$\beta<\lambda$ has been investigated. We now consider the situation
where $\beta\geq\lambda$, and study the bifurcation phenomena as
$\beta$ is varied. The value of $\gamma$ only changes the graph
quantitatively and hence it is not considered to be the
parameter. Moreover, $\rho$ is assumed to be positive. 

The behavior can be explained in the light of 
superstability of periodic orbits when border-collision
occurs. When $\beta=\lambda$ the points $\lambda,\infty$ form a
superstable 2-period orbit. As $\beta$ is increased, the product of
the derivatives approaches unity. At a point the period-2 orbit becomes
unstable giving way to a stable period-1 orbit. This is a standard
period-doubling bifurcation when $\beta$ is reduced. The nature of the
map (Fig.~\ref{infmap}) also suggests that period-1 orbit should
be stable at very high values of $\beta$. This behavior is shown in
the bifurcation diagram of Fig.~\ref{betabifurcation}.

However, if the initial iterate starts on the left-hand
side of the map and remains forever on the left-hand side, then the
behavior as shown in Fig.~\ref{betabifurcation} is not exhibited.If $0\leq\gamma+\frac{\alpha}{\lambda^2}\leq 1$ 
then the origin forms a stable fixed point and iterates starting on the left hand
side of the map converge to the origin.

\section{Scaling in the period-adding cascade}

\begin{figure}[thb]
 \centering \includegraphics[width=4in]{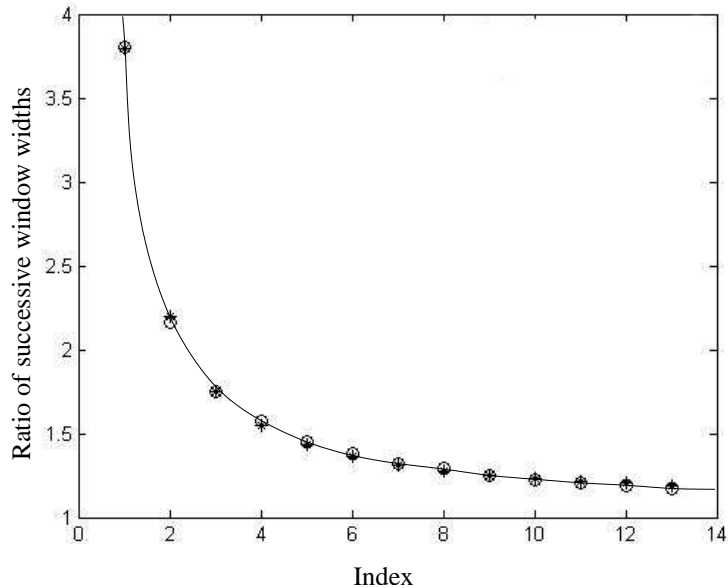}\\
  \caption{The ratios of the widths between the successive
    appearance of periodic windows. $*$ indicates the numerically
    determined values of the parameter ratios, and $\circ$ indicates
    those obtained from (\ref{eqn:fit}). }\label{fig.curve}
\end{figure}

Figure~\ref{fig.bifurcation} reveals that successive periodic
windows have monotonically decreasing width, which suggests the
existence of a  Feigenbaum-type ratio. To check this, we obtained the
parameter values at the points of saddle-node bifurcation where the
periodic orbits come into existence. The ratio of the widths
of successive periodic windows  were taken.
Fig.~\ref{fig.curve} shows the graph of the ratios of the widths
versus the index.

Using standard curve-fitting technique, we fit the data into the curve
\begin{equation}
    y =ax^b +c.
    \label{eqn:fit}
\end{equation}
The obtained value of $c$ was $1.091 \pm .033$. Hence the ratio of
successive widths will tend to $c$.

\section{Conclusions}

In this work a piecewise smooth map was considered that has
magnitude as well as slope singularity at a borderline value. It
was found that the singularity condition, i.e., the state 
assuming infinte value, occurs both in chaotic mode as well as in
periodic mode. We found that periodic orbits containing a point at
infinity occur are the points of border collision, at which these
orbits are superstable.  The system exhibits
period-adding cascade with deminishing width of successive
periodic windows, and the ratio converges to a number $1.091 \pm
.033$. There are several border-collisions within each
periodic window.

%\bibliographystyle{elsart-num}

%\bibliography{/home/soumitro/sourcefiles/sb}

\end{document}